\documentclass[aps,prd,showpacs,floatfix,preprint]{revtex4}
\usepackage{amsmath,bm}
\usepackage{graphicx}
\usepackage{epsfig}
\begin{document}

\title{Role of strongly magnetized crusts in torsional shear modes 
of magnetars}

\author{Rana Nandi$^{1}$, Debarati Chatterjee$^{2}$ and Debades Bandyopadhyay$^{1}$}
\affiliation{$^{1}$Astroparticle Physics and Cosmology Division,
Saha Institute of Nuclear Physics, 1/AF Bidhannagar, 
Kolkata-700064, India}
\affiliation{$^{2}$Institut f\"ur Theoretische Physik, Karl Ruprecht 
Universit\"at, Philosophenweg 16, 69120 Heidelberg, Germany}

\begin{abstract}
We study the influence of magnetised crusts on torsional shear mode 
oscillations of magnetars. In this context, we employ magnetised crusts whose 
ground state properties are affected by Landau quantisation of electrons.
The shear modulus of magnetised crusts is enhanced in strong magnetic fields
$\geq 10^{17}$ G. Though we do not find any appreciable change in frequencies 
of fundamental torsional shear modes, frequencies of first overtones are 
significantly affected in strong magnetic fields. Furthermore, frequencies of 
torsional shear modes calculated with magnetised crusts are in good agreement 
with frequencies of observed quasi-periodic oscillations.

\end{abstract}

\maketitle

\section{Introduction}
Soft gamma repeaters (SGRs) are characterised by their sporadic and short bursts
of soft gamma rays. Luminosities in these bursts could reach as high as 
$\sim 10^{41}$ ergs s$^{-1}$. There are about 5 SGRs known observationally.   
Evidences of stronger emissions of gamma rays from SGRs were observed in 
several cases. These events are known as giant flares in which luminosities 
are $\sim 10^{44}-10^{46}$ ergs s$^{-1}$. So far three cases of giant flares 
were reported and those are SGR 0526-66, SGR 1900+14 and SGR 1806-20 
\cite{bar,isr,watt1,watt2}. In giant flares, the early part of the spectrum 
was dominated by hard flash of shorter duration followed by a softer decaying 
tail of a few hundreds of seconds. 

SGRs are very good candidates for magnetars which are neutron stars with very 
high surface magnetic fields $\sim 10^{15}$ G \cite{dun,dun98}.
Giant flares might be caused by the evolving magnetic field and its stress on 
the crust of magnetars. It was argued that starquakes associated with 
giant flares could excite Global Seismic Oscillations (GSOs)\cite{dun98}. 
Torsional 
shear modes of magnetars with lower excitation energies would be easily 
excited. In this case, oscillations are restored by the Coulomb forces of 
crustal ions. Furthermore, the shear modes have longer damping times.  
Quasi-periodic oscillations (QPOs) were found in the decaying tail of giant 
flares from the timing analysis of data \cite{isr,watt1,watt2}. 
These findings implied that QPOs might be torsional shear mode oscillations 
of magnetar crusts \cite{dun98}. Frequencies of the observed QPOs ranged from 
18 Hz to 1800 Hz.

It was noted from earlier theoretical models of QPOs that the observed 
frequencies in particular higher frequencies could be explained reasonably well 
using torsional shear oscillations of magnetar crusts 
\cite{watt1,dun98,piro,sota1,sam,stei}. On the
other hand, lower frequencies of observed QPOs might be connected to Alfv\'en 
modes of the fluid core. This makes the study of the oscillations of magnetar
crusts more difficult. There were attempts to explain frequencies of QPOs using
Alfv\'en oscillations of the fluid core without considering a crust 
\cite{sota2,col,cer}. The coupling of Alfv\'en oscillations of fluid core with
the shear mode oscillations in the solid crust due to strong magnetic fields in
magnetars was already studied by several groups \cite{lev1,gla,lev2,le1,le2}.
It was argued that torsional shear modes of the crust might appear in GSOs and
explain frequencies of observed QPOs for not very strong magnetic fields 
despite all these complex problems \cite{sota3}.
 
Nuclear physics of crusts plays an important role on torsional shear modes
of magnetar crusts. In particular, the effects of the nuclear symmetry energy
on the shear mode frequencies were investigated recently \cite{stei}. It may be
worth noting here that torsional shear mode frequencies are sensitive to the 
shear modulus of neutron star crusts. Furthermore, the shear modulus 
strongly depends on the composition of neutron star crusts. Earlier studies of 
torsional shear mode oscillations exploited only non-magnetic crusts. Magnetic 
fields in magnetars may influence the ground state properties of neutron star 
crusts. Recently, we have investigated the influence of Landau quantisation of
electrons on the compositions and equations of state (EoS) of outer and inner 
crusts and obtained appreciable changes in those properties \cite{rana1,rana2}.
This, in turn, might influence the shear modulus of crusts and torsional shear 
mode oscillations in magnetars. This motivates us to study torsional shear mode
oscillations of magnetars using magnetised crusts.

We organise the paper in the following way. We describe models for 
torsional shear mode oscillations, shear modulus and compositions and EoS of 
magnetised crusts in Sec. II.
Results of this calculation are discussed in Sec. III. Section IV gives the
summary and conclusions.

\section{Formalism}
Earlier calculations of torsional shear mode oscillations were performed in 
Newtonian gravity \cite{dun98,piro,carr,mcd} as well as in general relativity 
\cite{sota1,sota2,sota3,kip,mes} with and without magnetic fields. In many of 
these calculations, the magnetised crust was decoupled from the fluid core.

As we are interested in the effects of magnetised crusts on torsional shear
mode frequencies, we consider a free slip between the crust and core.
Here we calculate torsional shear mode frequencies following the model
of Refs.\cite{sota1,mes}. In this case, we study torsional shear mode 
oscillations of spherical and non-rotating relativistic stars in the 
presence of a dipole magnetic field. We neglect the deformation in the 
equilibrium star due to magnetic fields $\sim 10^{16}$ G because its effects 
would be smaller than the direct influence of magnetic fields on shear mode 
frequencies. Therefore, we assume the magnetised star to be spherically 
symmetric. The metric used to determine equilibrium stellar models has the form,
\begin{equation}
ds^2 = - e^{2\Phi} dt^2 + e^{2\Lambda} dr^2 + r^2 \left( d{\theta}^2 + sin^2{\theta} d{\phi}^2 \right)~.
\end{equation}
The equilibrium models are obtained by solving the Tolman-Oppenheimer-Volkoff
(TOV) equation with a perfect fluid EoS. The stress-energy tensor of a 
magnetised relativistic star in equilibrium is written as
\begin{equation} 
T^{\mu\nu} = T^{\mu\nu}_M + T^{\mu\nu}_{EM},~
\label{stress}
\end{equation}
where the stress-energy tensor of perfect fluid is
\begin{equation}
T^{\mu\nu}_M = (\epsilon + P) u^{\mu} u^{\nu} + P g^{\mu\nu}~,
\end{equation} 
and
that of the electromagnetic field is
\begin{equation}
T^{\mu\nu}_{EM} = {\frac{1}{4\pi}}\left(B^2 u^{\mu}u^{\nu} + {\frac{1}{2}} 
g^{\mu\nu} B^2 - B^{\mu}B^{\nu}\right)~,
\end{equation}
where the four velocity of the fluid element is $u^{\mu} = (e^{-\Phi},0,0,0)$
and $u^{\mu}u_{\mu}=-1$.

Here we consider an axisymmetric poloidal magnetic field generated by four 
current $J_{\mu} = (0,0,0,J_{\Phi})$. After expanding the four-potential 
into vector spherical harmonics as 
$A_{\mu} = a_{\ell_m}(r) sin{\theta} {\partial_{\theta}}P_{\ell_m}(cos{\theta})$
and similarly for $J_{\mu}$, it follows from Maxwell's equations
for a dipole magnetic field i.e. $\ell_m=1$ 
\begin{equation}
   e^{-2\Lambda}\frac{d^2a_1}{dr^2} + (\Phi' - \Lambda')e^{-2\Lambda}
   \frac{da_1}{dr} -\frac{2 a_1}{r^2} = - 4\pi j_1~, 
     \label{a1}
\end{equation}
where prime denotes derivative with respect to $r$. The exterior solution 
for $a_1$ is given by \cite{sota1,mes}
  \begin{equation}
   a_1^{\rm{(ex)}} = -\frac{3\mu_br^2}{8M^3}\left[\ln\left(1-\frac{2M}{r}\right) + \frac{2M}{r}
     + \frac{2M^2}{r^2}\right],
\end{equation}
where $\mu_b$ is the magnetic dipole moment for an observer at infinity and $M$
is the mass of the star. The interior solution can be obtained by integrating 
Eq.(\ref{a1}) assuming a current distribution $j_1$ as was done by 
Ref.\cite{sota1}. 
The components of the magnetic field are 
\begin{eqnarray}
B_r &=& \frac{2 e^{\Lambda} cos{\theta}}{r^2}a_1\nonumber\\
B_{\theta} &=& - {e^{-\Lambda} sin{\theta}} \frac{da_1}{dr}~.
\end{eqnarray}

The perturbed equations describing shear mode oscillations are obtained by 
linearising the equations of motion of the fluid and the magnetic induction 
equation \cite{sota1,mes}. Torsional shear modes are incompressible and do not 
result in any appreciable density perturbation in equilibrium stars. 
Consequently, one may adopt the relativistic Cowling approximation 
and neglect metric perturbations $\delta g_{\mu\nu}$=0 \cite{mac83}. 
Torsional shear modes are the results of material velocity oscillations. 
These modes are odd parity (or axial type) modes. We consider axial type 
perturbation in the four velocity and
the relevant perturbed matter quantity 
is the $\phi$-component of the perturbed four velocity 
$\partial {u^{\phi}}$ \cite{sota1}
\begin{equation}
\partial {u^{\phi}} = e^{-\Phi} \partial_t {\cal{Y}}(t,r) {\frac{1}{sin{\theta}}}{\partial_{\theta}} P_l(cos{\theta})~,
\end{equation}
where $\partial_t$ and $\partial_{\theta}$ correspond to partial derivatives
with respect to time and $\theta$, respectively, $P_l(cos{\theta})$ is 
the Legendre polynomial of order $l$ and ${\cal{Y}}(t,r)$ is the angular 
displacement of the matter. It is to be noted that the radial and angular
variations of azimuthal displacement of stellar matter lead to
shears of the crystal lattice in neutron star crusts which are described by
the shear tensor $S_{\mu\nu}$ \cite{kip}. Further, the shear stress tensor is 
given by $T_{\mu\nu} = - 2 {\mu} S_{\mu\nu}$, where $\mu$ is the isotropic 
shear modulus. The linearised equations of motion includes the contribution of 
$\delta T_{\mu\nu}$ \cite{sota1}. 
 
Assuming a harmonic time dependence for ${\cal Y}(t,r) = e^{i\omega t} {\cal Y}
(r)$ and neglecting $l\pm2$ terms, one obtains the eigenvalue equation for
the mode frequency \cite{sota1}
{\scriptsize \begin{eqnarray}
 \Bigg[\mu + (1 + 2 \lambda_1)\frac{{a_1}^2}{\pi r^4}\Bigg]{\cal Y}''
    &+& \left\{\left(\frac{4}{r} + \Phi' - \Lambda'\right)\mu + \mu'
     + (1 + 2\lambda_1)\frac{a_1}{\pi r^4}\left[\left(\Phi'
- \Lambda'\right)a_1
+ 2{a_1}'\right]\right\}{\cal Y}'\nonumber \\
    &+& \Bigg\{\left[\left(\epsilon + p +
(1 +2\lambda_1)\frac{{a_1}^2}{\pi r^4}
\right)e^{2\Lambda}
     - \frac{\lambda_1 {{a_1}'}^2}{2\pi r^2}\right]\omega^2 e^{-2\Phi}
\nonumber \\
    &-&(\lambda-2)\left(\frac{ \mu e^{2\Lambda}}{r^2}
- \frac{\lambda_1{{a_1}'}^2}{2\pi r^4}\right)
     + \frac{(2 + 5\lambda_1)a_1}{2\pi r^4}\left\{\left(\Phi'
- \Lambda'\right){a_1}' + {a_1}''\right\}
     \Bigg\}{\cal Y}\nonumber\\  &=& 0~, 
\label{eigen}
\end{eqnarray}}
where $\lambda = \ell (\ell + 1)$ and 
$\lambda_1 = - \ell (\ell + 1)/(2\ell - 1)(2\ell + 3)$.
Equation 
(\ref{eigen}) reduces to the non-magnetic case when we put $a_1=0$ \cite{sota3}.
With suitable choice of new variables, Eq.(\ref{eigen}) results in a system of
first order ordinary differential equations \cite{sota1}. As torsional shear
modes are confined to the crust in our calculation, we impose a zero traction
boundary condition at the interface between the core and the crust as well
as the zero torque condition at the surface \cite{sota1,sota3}. These 
conditions imply ${\cal Y}' =0$ at surface ($r=R$) of the star and the 
interface ($r=R_c$) of the crust and core. Finally, we estimate 
eigenfrequencies by solving two first order differential equations.

The knowledge of the shear modulus of the magnetised crusts is an 
important input in the eigenvalue equation (Eq.(\ref{eigen})) for the 
shear mode calculation. Here we adopt the following expression of the shear 
modulus at zero temperature \cite{ichi,stroh} 
\begin{equation}
\mu = 0.1194 \frac{n_i (Ze)^2}{a}~,
\label{shr}
\end{equation}
where $a = 3/(4 \pi n_i)$, $Z$ is the atomic number of a nucleus and $n_i$ is
the ion density. This form of the shear modulus was obtained by assuming a bcc 
lattice and performing directional averages \cite{han}. Further the dependence 
of the shear modulus on temperature was also investigated with the Monte Carlo 
sampling technique by Strohmayer et al. \cite{stroh}. The composition and 
equation of state of neutron crusts are essential ingredients for the 
calculation of the shear modulus as it is evident from Eq.(\ref{shr}). In the 
following paragraphs, we describe the ground state properties in outer and 
inner crusts in presence of strong magnetic fields. 

The outer crust is composed of nuclei immersed in a uniform background of a 
noninteracting electron gas. Neutrons start coming out of nuclei when the
neutron drip point is reached. This is the beginning of the inner crust where
nuclei are placed both in free neutrons as well as electrons. Nuclei are 
arranged in a bcc lattice in neutron star crusts. The Wigner-Seitz 
approximation is adopted in this case. Each lattice volume is replaced by a
spherical cell with one nucleus at its center. The cell contains equal numbers 
of protons and electrons so that it is charge neutral. Moreover, there is no 
Coulomb interaction among cells. We obtain an equilibrium nucleus in the outer
crust minimising the Gibbs free energy per particle at a fixed pressure varying
mass and atomic numbers \cite{bps}
\begin{eqnarray}
g = \frac{E_{tot} + P}{n_b},
\label{out}
\end{eqnarray}
where $n_b$ is the baryon density and the energy density 
$E_{tot} = n_N (W_N + W_L) + \varepsilon_e$ includes contributions from 
the energy of the nucleus ($W_N$), lattice energy ($W_L$) of the cell involving
the finite size effects and free electron gas ($\varepsilon_e$) 
\cite{rana1}. Similarly, the total pressure is given by sum of the pressure 
due to the lattice and that of the electron gas. We use experimental nuclear 
masses from the atomic mass table \cite{audi03} whenever it is available. 
Otherwise, the theoretical extrapolation is adopted in this calculation 
\cite{moller95}.

We describe the ground state properties of matter of the inner crust using the
Thomas-Fermi model. The spherical cell that contains
neutrons and protons does not define a nucleus. We adopt the procedure of 
Bonche, Levit and Vautherin to subtract the free neutron gas of the cell 
and obtain the nucleus \cite{bon1,bon2}. Under this assumption, the 
thermodynamic potential ($\Omega_{N}$) of the nucleus is obtained as 
\cite{bon1,bon2} 
\begin{equation}
\Omega_{N} = \Omega_{NG} - \Omega_{G},
\end{equation}
where $\Omega_{NG}$ is the thermodynamical potential for the nucleus plus the
free neutron gas and $\Omega_{G}$ is that of the free neutron gas. The
thermodynamical potential is defined as 
\begin{equation}
\Omega = {\cal{F}} - \sum_{q=n,p} \mu_q n_q~,
\end{equation}
where $\cal{F}$, $\mu_q$ and $n_q$ are the free energy density, baryon
chemical potential and number density, respectively. The free energy is given
by
\begin{equation}
{\cal{F}}(n_q,Y_p) = \int [{\cal{H}} + \varepsilon_c + \varepsilon_e] d{\bf r}~,
\label{free}
\end{equation}
where $Y_p$ is the proton fraction. Equation (\ref{free}) includes 
contributions from the nuclear energy density ($\cal{H}$), Coulomb energy 
density ($\varepsilon_c$) and energy density of free electron gas 
($\varepsilon_e$) \cite{sil,rana2}. The nuclear energy density is calculated 
using the Skyrme nucleon-nucleon interaction \cite{kri,bra,sto}. We obtain the
density profiles of neutrons and protons for the nucleus plus neutron gas as
well as neutron gas phases by minimising thermodynamical potentials and 
hence calculate mass and atomic numbers of nuclei of the inner crust 
\cite{rana2}. 

Now we focus on neutron star crusts in strongly quantising magnetic fields. In
particular, we showed earlier that the Landau quantisation of electrons strongly
influenced ground state properties of neutron star crusts in strong magnetic
fields $\sim 10^{17}$ G \cite{rana1,rana2}. Energy and number densities of 
electrons are affected by the phase space modifications due to Landau 
quantisation of electrons. We immediately note that the electron energy density
in Eqs.(\ref{out}) and (\ref{free}) are modified in presence of magnetic 
fields \cite{rana1,rana2}. Further, we have to take into account the change
in the average electron chemical potential due to magnetic fields in the outer
and inner crusts. It is to be noted that protons are only influenced by 
magnetic fields through the charge neutrality condition.

\section{Results and Discussions}
We investigated the composition and EoS of ground state matter in
neutron star crusts in strong magnetic fields \cite{rana1,rana2}. We noted that
the electron number density in the outer crust was enhanced compared with 
the field free case when a few Landau levels were populated for magnetic
fields $> B=4.414 \times 10^{16}$ G \cite{rana1}. It was observed that this 
enhancement grew stronger when only the zeroth Landau level was populated at a
magnetic field strength of 4.414 $\times 10^{17}$ G. Consequently, we found
modifications in the sequence of equilibrium nuclei which was obtained by 
minimising the Gibbs free energy per nucleon of Eq.(\ref{out}). It was noted
that some new nuclei such as $^{88}_{38}$Sr and $^{128}_{46}$Pd appeared
and some nuclei such as $^{66}_{28}$Ni and $^{78}_{28}$Ni disappeared in a 
magnetic field
of $B=4.414 \times 10^{16}$ G \cite{rana1} when we compared this with the zero 
field case.
The maximum density upto which an equilibrium nucleus would exist, increased
as the strength of magnetic field increased. This implied that the neutron drip
point was
shifted to higher density in presence of a strong magnetic field with 
respect to the field free case \cite{rana1}. We performed the calculation of
the inner crust using the SkM nucleon-nucleon interaction \cite{rana2,sil}.
In this case too, we calculated the equilibrium nucleus at each density point.
Like the outer crust in strong magnetic fields, the electron number
density was enhanced due to the electron population in the zero Landau level 
for
magnetic fields $\geq 10^{17}$ G which , in turn, led to a large proton fraction
because of charge neutrality. For magnetic fields $>10^{17}$ G, equilibrium
nuclei with larger mass and atomic numbers were found to exist in the crust
\cite{rana2}. Furthermore,
the free energy per nucleon of the nuclear system was reduced in magnetic 
fields compared with the corresponding case without a magnetic field. 

We calculate the shear modulus using Eq.({\ref{shr}) and the above mentioned 
models of magnetised crusts. For this purpose, we have to know the profiles of 
pressure, energy density and shear modulus in a neutron star. Those profiles
are obtained by solving the TOV equation. In this context, we construct an EoS 
of dense nuclear matter in strong magnetic fields in neutron star core using a 
relativistic mean field model with the GM1 
parameter set as described in Ref.\cite{cbp,sb,Gle91}. This EoS of dense 
nuclear matter
is matched with the EoS of the crust and used in the TOV equation. Figure 1 
displays the shear modulus as a function of mass density for a neutron star of
1.4 $M_{\odot}$. 
The shear modulus (dahsed line) corresponding to $B_{*}= B/B_c=10^3$ where 
$B_c=4.414 \times 10^{13}$ G, does not show any appreciable change from that of
the zero field (dotted line) because of large numbers of Landau levels are
populated in this field. As the field strength is increased, less numbers
of Landau levels are populated. For $B_{*} = 10^4$ i.e. 
$4.414 \times 10^{17}$ G, the shear modulus is enhanced due to the
population of all electrons in the zeroth Landau level. In all three cases, the 
shear modulus increases with mass density well before the crust-core interface.
The shear modulus and shear speed $v_s=(\mu/\rho)^{1/2}$ are extrapolated to the
zero value at the crust-core interface for magnetised as well as non-magnetised
crusts. It was noted that the shear 
modulus was found to decrease smoothly to zero with increasing density when the 
pasta phase was considered in the crust in absence of
magnetic fields \cite{sota3}. 
We generate a profile of the shear modulus as a function of
radial distance in a neutron star for calculating frequencies of torsional
shear modes.
      
Now we study the dependence of torsional shear mode frequencies on the 
compositions and the shear modulus of magnetised crusts. Earlier all 
calculations were performed using non-magnetic neutron star crusts. Here we
exploit models of non-magnetic as well as magnetic crusts which are already 
described in this section. We consider torsional shear modes of a neutron star 
with 1.4 $M_{\odot}$. Frequencies of fundamental ($n=0$, $\ell=2$) torsional 
shear modes 
are plotted with magnetic fields in Fig. 2. It is observed that the frequency 
increases
with magnetic field. However, there is no differences between our results with
and without magnetised crusts. Figure 3 shows frequencies of torsional shear
modes corresponding to $n=0$ plotted as a function of $\ell$ values for a
1.4 $M_{\odot}$ neutron star and magnetic field $B_{*}=10^3$ i.e. $4.414 \times
10^{16}$ G. Again we calculate frequencies using the non-magnetic and magnetic
crusts in this case. The frequency increases with higher $\ell$ values, but we
can not distinguish between the results of non-magnetic and magnetic crusts. 

We continue our investigation on frequencies of first overtones ($n=1$) of
torsional shear modes in presence of a magnetic field. Frequencies of first
overtones are plotted with $\ell$ values for a neutron star of 1.4 $M_{\odot}$
and magnetic field $B_{*}=10^3$ i.e. $4.414 \times 10^{16}$ G in Fig. 4. In 
this case,
the results (dashed line) obtained with magnetised crusts deviate from those of
non-magnetised crusts (solid line). We obtain qualitatively similar results for
$B_{*}=10^4$. It was noted that the ratio of the crust
thickness to the radius of a neutron star could be obtained from the 
frequencies of overtones \cite{sota1}. Results of Fig. 4 point to the fact
that the ratio of the crust thickness to the radius might be influenced by
strong magnetic fields. 

The dependence of frequencies of the fundamental mode and higher harmonics
on neutron star masses is demonstrated in Fig. 5. Here the frequencies 
corresponding to $n=0$ and $\ell = 2,3,4$ are shown as a function of neutron 
star masses for a magnetic field $B=8 \times 10^{14}$ G. For each case, 
frequencies of torsional shear modes decrease with increasing mass, whereas
higher $\ell$ values lead to higher frequencies. Frequencies of observed QPOs 
might put a strong constraint on the EoS if masses of neutron stars are known
accurately.  

Finally, we compare the calculated frequencies of torsional shear modes with
frequencies of observed QPOs. This comparison is shown in Table 1. Our results
are obtained using the magnetised crusts of a 1.2 $M_{\odot}$ neutron star and 
estimated magnetic fields $B=8 \times 10^{14}$ G corresponding to SGR 1806-20 
and 1.4 $M_{\odot}$ neutron star and $B= 4 \times 10^{14}$ G for SGR 1900+14 
\cite{kou,sota4,hur}. For 
SGR 1806-20, calculated frequencies 93 Hz corresponding to $n=0$ and 
$\ell=12$ and above are in very good agreement with the observed frequencies. 
However, lower frequencies 18, 26 and 29 Hz can not be explained with our 
calculation because those are so close that they can not be matched using 
different harmonics of $n=0$ mode.  On the other hand,
the overall agreement of calculated frequencies with the observed frequencies 
of SGR 1900+14 is excellent.  
      
\section{Summary and Conclusions}
We have estimated frequencies of torsional shear modes of magnetars assuming
a dipole magnetic field configuration. Frequencies are computed using our 
models of magnetised crusts. The shear modulus of magnetised crusts is found to
be enhanced in strong magnetic fields $\sim 4.414 \times 10^{17}$ G because 
electrons populate the zeroth Landau level. It is observed that frequencies of 
fundamental ($n=0$, $\ell=2$) torsional shear modes are not sensitive to this 
enhancement in the shear modulus in strong magnetic fields. On the other hand, 
frequencies of first overtones ($n=1$) of torsional shear modes in presence of 
strongly quantising magnetic fields are distinctly different from those of the 
field free case. Consequently, this might impact the ratio of the crust 
thickness to the radius of a magnetar. We have compared our results with 
frequencies of observed QPOs and found good agreement. Observed frequencies
could constrain the EoS of magnetised neutron star crusts if masses of neutron
stars are known.

\newpage
\begin{table}
\caption {Frequencies of torsional shear modes are recorded here.
Frequencies obtained in the calculation using the magnetised crusts 
\cite{rana1,rana2} are compared with observed frequencies in SGR 1806-20 and 
SGR 1900+14. Our results are calculated using a 1.2 $M_{\odot}$ neutron star 
and magnetic field $8 \times 10^{14}$ G for SGR 1806-20 and a 1.4 $M_{\odot}$ 
neutron star and magnetic field $4 \times 10^{14}$ G for SGR 1900+14.}
\begin{center}
\vspace{2cm}
 \begin{tabular}{|c|c|c|c|c|c|c|c|}
 \hline
 \multicolumn{4}{|c|}{SGR 1806-20}&\multicolumn{4}{|c|}{SGR 1900+14}\\
 \cline{1-8}
 Observed & Calculated & & & Observed & Calculated & & \\
 Frequency & Frequency & n & $\ell$&Frequency & Frequency & n & $\ell$\\
 (Hz) & (Hz) &&&(Hz) & (Hz) &&\\
 \hline\hline
 18 & 15 & 0 & 2& 28 & 28 & 0 & 4\\
 26 & 24 & 0 & 3& 54 & 55 & 0 & 8\\
 29 & 32 & 0 & 4& 84 & 82 & 0 &12\\
 93 & 93 & 0 &12&155 &154 & 0 &23\\
 150 & 151 & 0 & 20&&&&\\
 626 & 626 & 1 & 29&&&&\\
 1838 & 1834 & 4 & 2&&&&\\
 \hline
 \end{tabular}
\end{center}
\end{table}
\newpage
\vspace{-2cm}

{\centerline{
\epsfxsize=12cm
\epsfysize=14cm
\epsffile{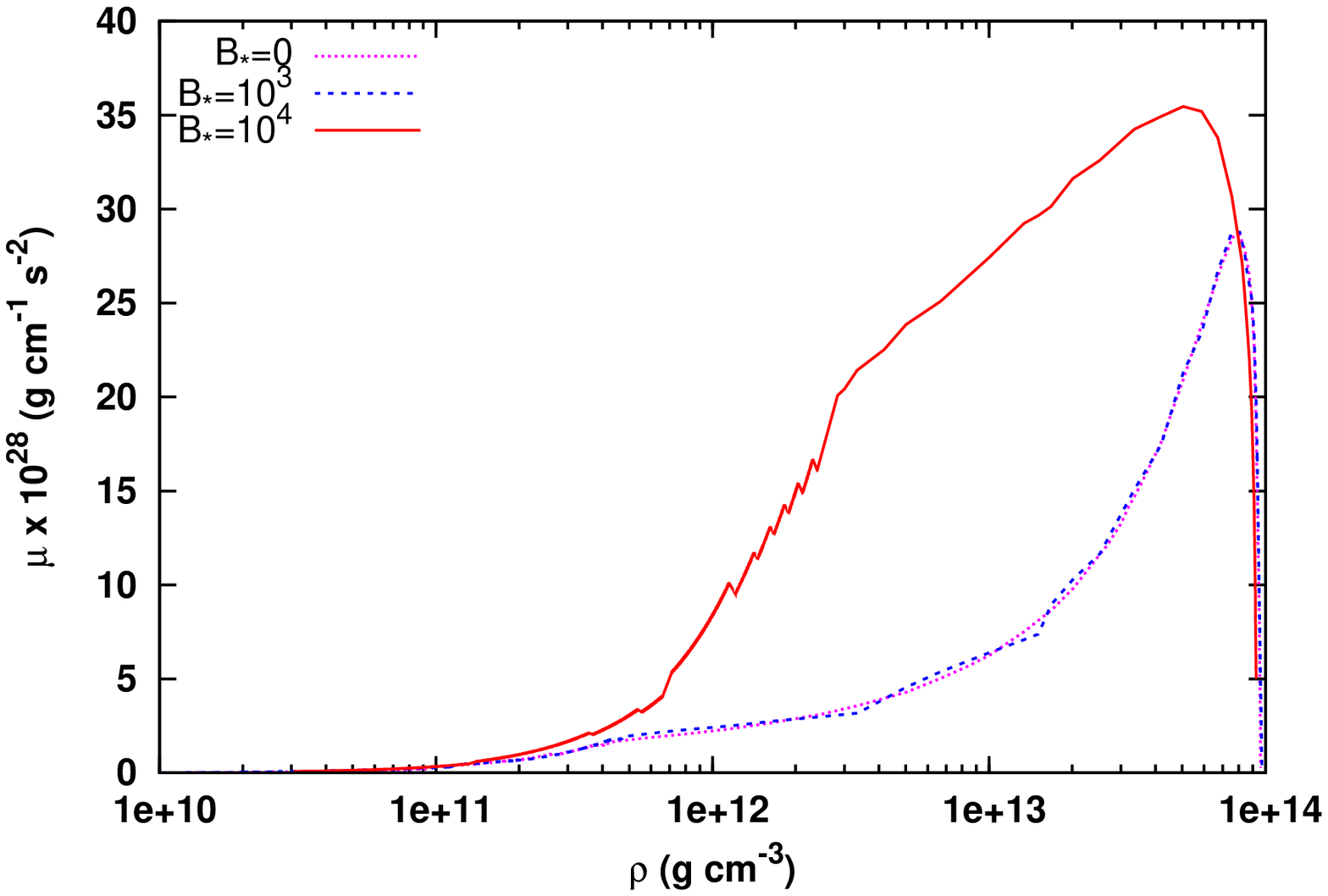}
}}

\vspace{4.0cm}

\noindent{\small{
FIG. 1. Shear modulus is plotted as a function of mass density for a neutron 
star of 1.4 $M_{\odot}$ and different magnetic fields.}}

\newpage
\vspace{-2cm}

{\centerline{
\epsfxsize=12cm
\epsfysize=14cm
\epsffile{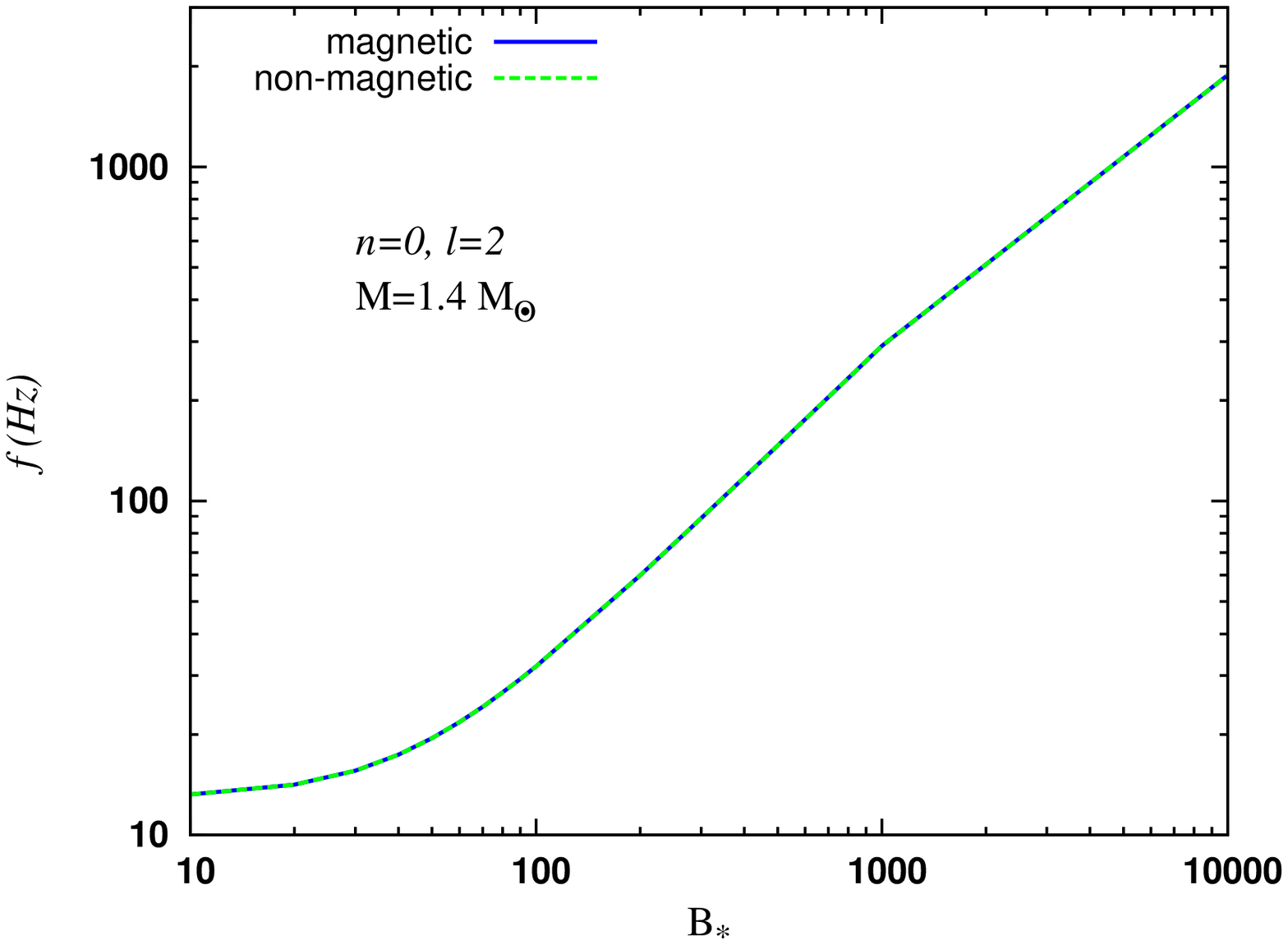}
}}

\vspace{1.0cm}

\noindent{\small{Fig. 2.  Frequency of fundamental ($n=0$, $\ell=2$) torsional
shear mode for  a neutron star of 1.4 $M_{\odot}$ is shown as a function of 
magnetic field $B_{*}=B/B_c$ where $B_{c}=4.414 \times 10^{13}$ G. Results of 
our calculations with and without magnetised crusts are shown here.}}

\newpage 
\vspace{-2.0cm}

{\centerline{
\epsfxsize=12cm
\epsfysize=14cm
\epsffile{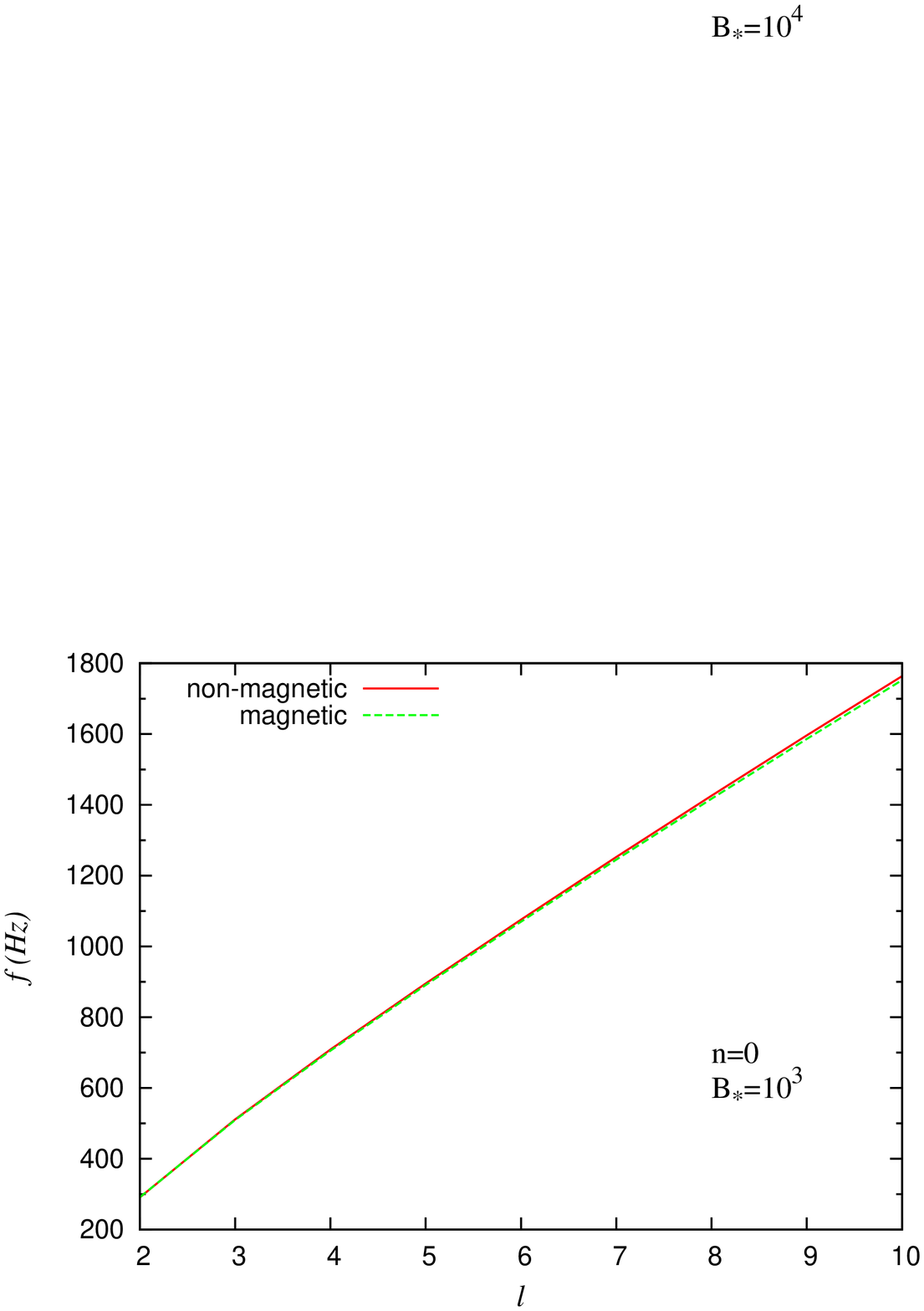}
}}

\vspace{1.0cm}

\noindent{\small{Fig. 3. Frequency of $n=0$ torsional shear mode as a function
of $\ell$ values with and without magnetic crusts of a 1.4 $M_{\odot}$ neutron
star for magnetic field $B_{*}=10^3$.}}

\newpage 
\vspace{-2cm}

{\centerline{
\epsfxsize=12cm
\epsfysize=14cm
\epsffile{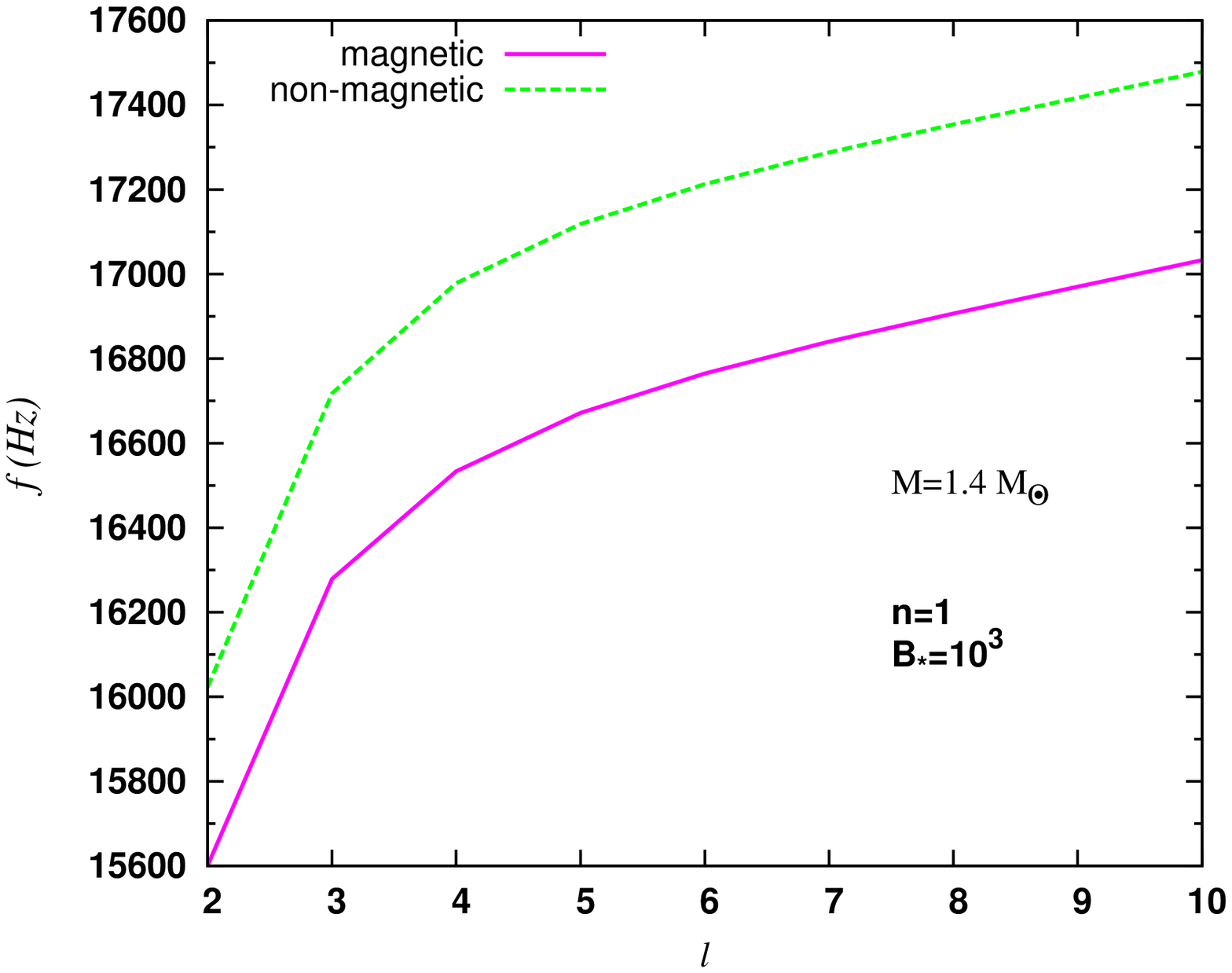}
}}

\vspace{4.0cm}

\noindent{\small{ Fig. 4 Frequency of first overtone ($n=1$) of torsional shear
mode of $\ell$ values with and without magnetic crusts of a 1.4 $M_{\odot}$ 
neutron star for magnetic field $B_{*}=10^3$.}}

\newpage 
\vspace{-2cm}

{\centerline{
\epsfxsize=12cm
\epsfysize=14cm
\epsffile{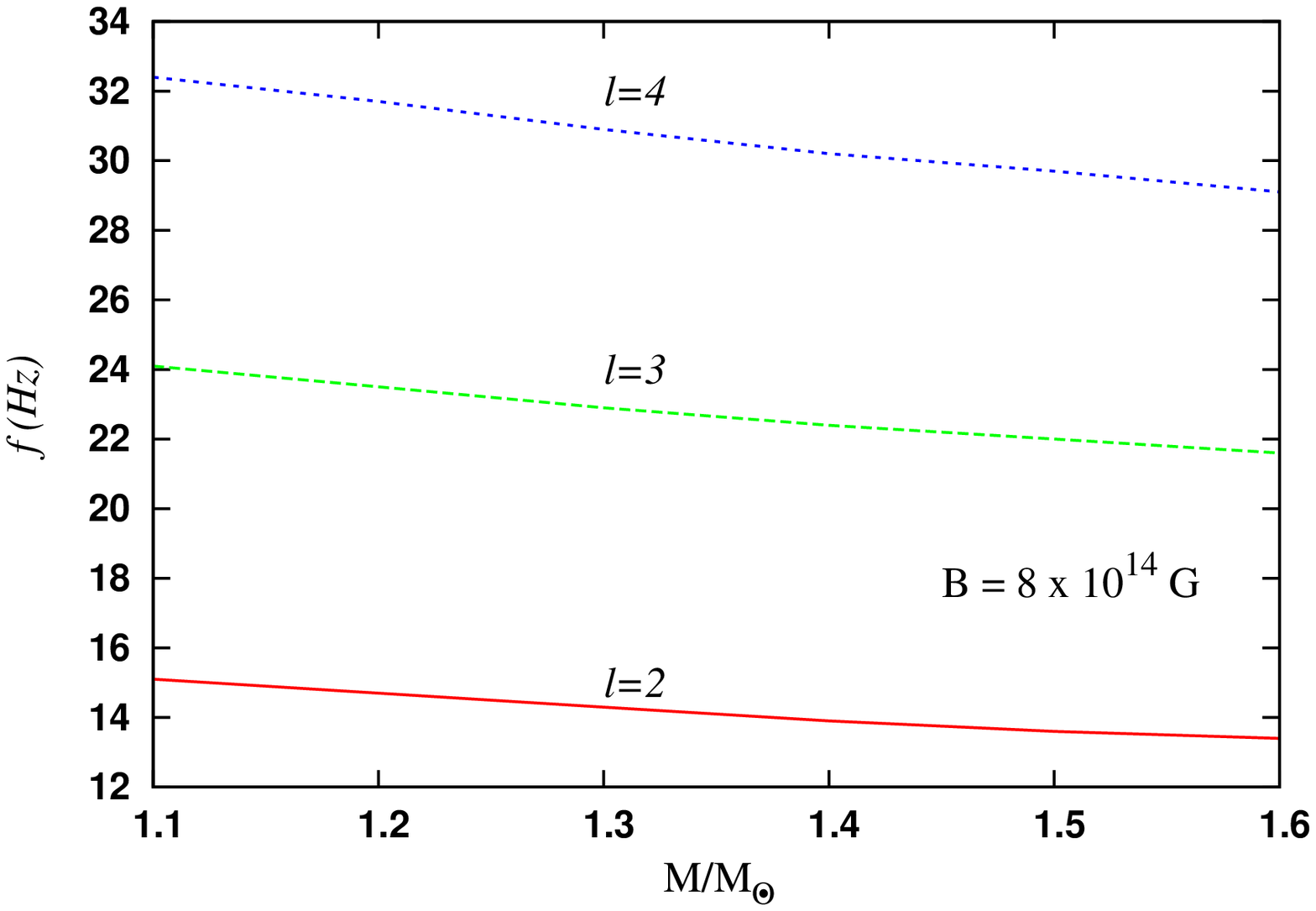}
}}

\vspace{4.0cm}

\noindent{\small{
FIG. 5. Frequency of torsional shear mode corresponding to $n=0$ and 
$\ell=2,3,4$ as a function of neutron star mass for a magnetic field 
$B = 8 \times 10^{14}$ G.}}

\end{document}